\documentclass{article}
\usepackage{ismir,cite}
\usepackage{aliascnt}
\usepackage{tabu}
\usepackage{amssymb,amsmath,mathtools,amsthm}
\usepackage{booktabs}
\usepackage{multirow}
\usepackage[english]{babel}

\usepackage{epstopdf}
\usepackage{gensymb}
\usepackage{siunitx}
\usepackage[draft]{todonotes}   

\usepackage{times}
\usepackage{url}
\usepackage{color}
\usepackage{multirow}
\usepackage{multicol}
\usepackage{graphicx}
\usepackage{caption}
\usepackage{tikz}
\usetikzlibrary{calc,tikzmark}
\usepackage[shortcuts]{extdash}
\usepackage{blindtext}
\usepackage[ruled,norelsize,linesnumbered]{algorithm2e}

\newcommand{\comment}[1]{\textcolor{blue}{}}
\newcommand{\mg}[1]{\textcolor{red}{}}

\makeatletter
\def\inmod#1{\allowbreak\mkern5mu{\operator@font mod}\,\,#1}
\makeatother

\title{SampleMatch: Drum Sample Retrieval by Musical Context}

\oneauthor
{Stefan Lattner}
{Sony Computer Science Laboratories (CSL), Paris, France \\ {\tt stefan.lattner@sony.com}}
\def\authorname{Stefan Lattner}

\begin{document}

\maketitle

\begin{abstract}
Modern digital music production typically involves combining numerous acoustic elements to compile a piece of music. Important types of such elements are \emph{drum samples}, which determine the characteristics of the percussive components of the piece. Artists must use their aesthetic judgement to assess whether a given drum sample fits the current musical context. However, selecting drum samples from a potentially large library is tedious and may interrupt the creative flow. In this work, we explore the automatic drum sample retrieval based on aesthetic principles learned from data. As a result, artists can rank the samples in their library by fit to some musical context at different stages of the production process (i.e., by fit to incomplete song mixtures). To this end, we use contrastive learning to maximize the score of drum samples originating from the same song as the mixture. We conduct a listening test to determine whether the human ratings match the automatic scoring function. We also perform objective quantitative analyses to evaluate the efficacy of our approach.
\end{abstract}

\section{Introduction}\label{sec:introduction}
Machine-learning (ML) powered tools are increasingly finding their way into modern music production \cite{TISMIR_CSL}. Commercial tools already use machine learning for different applications, like mixing/mastering\footnote{\url{https://www.izotope.com/}}or data visualization.\footnote{\url{https://algonaut.audio/}} In public perception, ML and artificial intelligence (AI) is often associated with \emph{content generation}. Indeed, also in music production, generative AI models are about to transition from research labs to the studio, in the form of tools that integrate seamlessly into the artist's workflow, like Magenta Studio\footnote{\url{https://magenta.tensorflow.org/studio/}} and others \cite{DrumGAN_VST, DBLP:conf/iclr/EngelHGR20, NOTONO, DrumNet, PIA}.

Besides content generation, another frequent task in modern music production is \emph{content retrieval}, where commercial solutions (based on expert systems) also already exist.\footnote{\url{https://jamahook.com/}} In manual content retrieval (i.e., content selection), artists select ``loops'' (short, typically bar-aligned audio files of any content) or ``samples'' (usually shorter, single notes or drum hits) that fit the current musical context. Content selection is a creative act in its own right, but it is challenging to browse through potentially large loop or sample libraries while -- crucially -- keeping the current context in mind.

To support the selection process, we envision a retrieval system that learns aesthetic principles of matching drum samples to musical context from data. As a result, it is able to sort a drum sample library according to a \emph{score} it assigns to each sample. This lets users start their search with the most promising options first and may help turn a cumbersome curation process into a more artistic activity.

We use the cosine similarity in a learned latent space as a scoring function. Two encoders are trained with a contrastive loss to maximize the score between mixtures and drum samples originating from the same song. The encoders are trained with an electronic and acoustic data set of songs whose tracks are available separately. We perform a user study to evaluate if human listeners agree with our proposed scoring function. To that end, we test if participants prefer samples that obtain a high rating by the scoring function over randomly selected samples. Furthermore, we perform an ablation study to evaluate the different components of our method.
In addition, we want to understand better how the learned space is organized. In particular, we want to gain some insights into the relationship of drum samples that fit well in the same context. For that, we perform a correlation analysis between audio features of samples that are close in the learned latent space.

The paper is organized as follows. Related works are discussed in Section \ref{sec:related}, and Section \ref{sec:method} describes the used method. The used data and data preprocessing are described in Section \ref{sec:data}. Section \ref{sec:training} explaines how the model training is performed. In Section \ref{sec:experiments}, we introduce the performed experiments, including the objective evaluation, a user study and correlation analysis. We present and discuss the results in Section \ref{sec:results} and conclude in Section \ref{sec:conclusion}.

\section{Related Work}\label{sec:related}
Self-supervised learning has been a hot topic in recent years, with well-known works in the visual domain, like MoCo \cite{DBLP:conf/cvpr/He0WXG20} and SimCLR \cite{SimCLR} that adopt contrastive learning approaches, and BYOL \cite{DBLP:conf/nips/GrillSATRBDPGAP20}, or VICReg \cite{DBLP:journals/corr/abs-2105-04906} that aim to eliminate negative samples.
Examples for contrastive approaches in the speech domain are Wav2Vec \cite{DBLP:conf/interspeech/SchneiderBCA19}, Wav2Vec 2.0 \cite{DBLP:conf/nips/BaevskiZMA20}, as well as SpeechSimCLR \cite{DBLP:conf/interspeech/JiangLCZL21}.
Inspired by these works, self-supervised learning in non-speech (i.e., musical and environmental) audio was also introduced with contrastive approaches \cite{DBLP:conf/icassp/SaeedGZ21, DBLP:conf/ismir/SpijkervetB21}, Audio2Vec \cite{DBLP:journals/spl/TagliasacchiGQR20} and Wav2Vec for non-speech audio using a conformer architecture \cite{DBLP:journals/corr/abs-2110-07313}. An example for self-supervised learning for audio based on the BYOL method is \cite{DBLP:conf/ijcnn/NiizumiTOHK21}. Other audio-related contributions based on contrastive learning are multimodal contrastive learning (for audio and video) \cite{DBLP:journals/corr/abs-2104-12807}, as well as multi-format contrastive learning (between raw audio and spectrogram representations) \cite{DBLP:journals/corr/abs-2103-06508}. We adopt ideas from self-supervised learning methods mentioned above, using a dictionary look-up approach like in MoCo \cite{DBLP:conf/cvpr/He0WXG20}, we adopt the variance- and co-variance regularization of VICReg \cite{DBLP:journals/corr/abs-2105-04906}, and we use decoupled contrastive learning, as proposed in \cite{decoupledContrastive}.

Other works on drum sample retrieval involve the contrastive learning-based retrieval of single drum samples by their mixed versions \cite{kim2020drum} and retrieval by vocalization \cite{DBLP:journals/corr/abs-2204-04651, DBLP:conf/icassp/MehrabiCDS18}. Other academic works that tackle the task of matching artistic content based on \emph{aesthetic principles} are the Neural Loop Combiner \cite{DBLP:conf/ismir/ChenSY20}, and a stem mashup creation approach \cite{DBLP:conf/aaai/HuangWSSW21} (both based on contrastive learning). Other methods for computational mashup creation are based on optimization \cite{DBLP:conf/audio/BernardoB21} or expert systems \cite{DBLP:journals/taslp/DaviesHYG14}.

\section{Method}\label{sec:method}
\subsection{Scoring Function}
The goal of our proposed method is to estimate how well a given drum sample fits a specific musical context. This task can be modeled by contrastive learning as a dictionary look-up task, as described in \cite{DBLP:conf/cvpr/He0WXG20}. A musical context $\mathbf{q}_i \in \mathbb{R}^m$ acts as a query, and drum samples $\mathbf{k}_j \in \mathbb{R}^m$ as keys. We define a scoring function $s(\mathbf{q}_i, \mathbf{k}_j)$ that outputs a high score if the sample fits the query. To that end, we use two encoders $g_q$ and $g_k$ to produce a query encoding $\mathbf{u}_i = g_q(\mathbf{q}_i), \mathbf{u}_i \in \mathbb{R}^d$ and a sample encoding $\mathbf{v}_j = g_k(\mathbf{k}_j), \mathbf{v}_j \in \mathbb{R}^d$. Finally, we define the scoring function as the cosine similarity $\text{sim}(\cdot, \cdot)$ between the resulting encodings $s(\mathbf{q}_i, \mathbf{k}_j) = \text{sim}(\mathbf{u}_i, \mathbf{v}_j)$.
We train the encoders using a contrastive loss called NT-Xent in \cite{SimCLR}:
\begin{equation}\label{eq:loss}
\mathcal{X}(Z) = -\log \frac{\exp{(\text{sim}(\mathbf{u}_i, \mathbf{v}_j) / \tau})}{\sum_{l \neq j}{\exp{(\text{sim}(\mathbf{u}_i, \mathbf{v}_l) / \tau})}},
\end{equation}
where $\{\mathbf{u}_i, \mathbf{v}_j\}$ is a positive pair, $Z \in \mathbb{R}^{n \times d}$ are all representations of a training batch, $\tau$ is the temperature parameter, and we adopt the decoupled contrastive learning variant, that has shown to work better for smaller batch sizes, by removing the positive pair from the denominator (i.e., $l \neq j$)\cite{decoupledContrastive}.

\subsection{Regularizations}
Furthermore, we combine the contrastive loss with the variance and covariance regularization used in VICReg \cite{DBLP:journals/corr/abs-2105-04906}. The variance regularization term is defined as a hinge function that penalizes variances of latent features along the batch dimension that are smaller than 1 as

\begin{equation}\label{eq:var}
\mathcal{V}(Z) = \frac{1}{d} \sum_{j=1}^d{\max{(0, 1 - S(\mathbf{z}_{:,j}, \epsilon))}},
\end{equation}
where Python slicing notation is used, and $S$ is the regularized standard deviation
\begin{equation}
S(x, \epsilon) = \sqrt{\text{Var}(x) + \epsilon}.
\end{equation}

The covariance regularization penalizes non-zero off-diagonal entries in the covariance matrix of each batch, leading to a decorrelation of the latent dimensions:
\begin{equation}\label{eq:covar}
\mathcal{C}(Z) = \frac{1}{d} \sum_{i \neq j}{[C(Z)]_{i,j}^2},
\end{equation}
where $C$ is the covariance matrix
\begin{equation}
C(Z) = \frac{1}{d\!-\!1} \sum_{i=1}^d{(\mathbf{z}_{:,i} - \bar{z}_i)(\mathbf{z}_{:,i} - \bar{z}_i)^T}, \;  \bar{z}_i = \frac{1}{n} \sum_{j=1}^n{z_{j,i}}.
\end{equation}

Even though the cosine distance renders the norms of the network outputs irrelevant, they can still grow due to optimization dynamics. To keep the norms in a moderate range, we also add a norm regularization in form of a hinge function as

\begin{equation}
\mathcal{N}(Z) = -\frac{1}{n} \sum_{i=1}^{n} \min{(0, c - ||\mathbf{z}_i||)},
\end{equation}
where we fix $c=4$ in our experiments (because most norms are smaller than $4$ at initialization).
Putting all the above terms together (and weighting the regularization terms with factors $\gamma$, $\delta$ and $\eta$), yields the final loss

\begin{equation}
\mathcal{L}(Z) = \mathcal{X}(Z) + \gamma\mathcal{V}(Z) + \delta\mathcal{C}(Z) + \eta\mathcal{N}(Z).
\end{equation}

\section{Data}\label{sec:data}
We used a dataset of electronic music and pop/rock songs of $44.1$ kHz sample rate for training and evaluation. The electronic music portion consists of $4830$ so-called ``remix packs'', with durations ranging from several seconds to several minutes. Each remix pack consists of multiple audio tracks that contain the individual (mutually aligned) instruments (such as synth, bass, guitar, pad, strings, choir, brass, keyboard, vocals, and different percussion instruments).

The Pop/Rock dataset is a proprietary dataset of $885$ well-known rock/pop songs of the past decades, including artists such as Lady Gaga, Coldplay, Stevie Wonder, Lenny Kravitz, Metallica, AC/DC, and Red Hot Chili Peppers. The stems (guitar, vocals, bass guitar, keyboard, misc, and different percussion instruments) are available as individual audio tracks for each song.

From every percussion track in the dataset, we extract so-called ``one-shots'', single hits with the respective percussion instrument. To that end, we picked those samples where the subsequent onset is of maximal distance (to retain most of the percussion's potential reverb). From the pop/rock songs, we extract $5$ one-shots per percussion track, while from the electronic songs, we extract $1$ one-shot per percussion track (as most of the time they are all identical). Altogether, this results in $63042$ one-shot drum samples.

We categorize the extracted drum samples into $6$ categories which are \{kick, snare, hi-hat, ride, crash, toms\}. The assignment is based on the filenames of the stems the samples originate from, using keyword dictionaries of typical drum-type expressions (like ``hat'' for hi-hat or ``BD'' for kick drum). We split the dataset in train / eval / test set with proportions $0.85 / 0.05 / 0.1$ (on a track level, meaning that samples of one track do not spread over different sets). All reported results are computed on the test set.

\section{Training}\label{sec:training}
As an encoder, we use the EfficientNet-B4 \cite{efficientnet}, where we start training from weights that have been pre-trained on the ImageNet dataset. Even though ImageNet is far from the audio spectrum domain, it turned out that using ImageNet weights is crucial in our experiments (see Section \ref{sec:objective}, and it has also been shown in a previous study that cross-domain pre-training can be beneficial \cite{DBLP:conf/icpr/GuzhovRH020}).

The $1000$ output units are reduced to $256$ with a single linear layer. The EfficientNet expects a 2D input, which we provide by converting the audio signals into mel-scaled spectrograms with an STFT window length of 2048, a hop length of 512, and 128 resulting mel bins, considering the whole frequency range (fmax $=22050$). Also, we log-scale the values of the resulting mel spectrograms. We input the resulting spectrograms in each of the three RGB input channels of the EfficientNet (after normalization to the ImageNet color statistics).

A positive pair consists of an audio query and a corresponding drum sample. Given a drum sample, we select the stems of the corresponding song and remove the stem from which the drum sample was extracted. Then, we randomly choose at least 2, at most all of the remaining stems (the number of stems is uniformly sampled) and mix them on the fly during training (note that the mel spectrogram transformation is part of the model architecture using the \emph{nnAudio} library).\footnote{\url{https://github.com/KinWaiCheuk/nnAudio}} All audio inputs are of length 4 seconds. If a drum sample is shorter than that, it is zero-padded. To obtain a query, we are cutting a 4-second-long snippet from a random position of the mixture.

The encoders are trained by the ADAM optimizer, with a batch size of $190$, a learning rate of $3$e-$4$, and a weight decay factor of $3$e-$5$. The temperature parameter $\tau$ of the NT-Xent loss is set to $0.2$. The factors for the variance and covariance regularization terms $\gamma$ and $\delta$ are set to $1$, and the norm regularization factor $\eta$ is set to $10$.

We use some data augmentation on both the queries and the drum samples. First, Gaussian noise is added (up to $-12$ dB). Furthermore, we perform time-stretch with a ratio between $0.9$ and $1.15$ of the original tempo. Another augmentation is reducing the gain down to a minimum attenuation of $-6$ dB, and we perform a time shift of up to $800$ms to the right and $200$ms to the left (to make the models invariant to the exact onset position of one-shot audios). All these augmentations occur with a probability of $50\%$ for each instance.

\section{Experiments}\label{sec:experiments}
In this section, we introduce the conducted experiments to validate our method including objective metrics (Section \ref{sec:metrics}), the user study (Section \ref{sec:user_study_method}), and we describe the method of the correlation analysis (Section \ref{sec:correlation_method}).

\subsection{Objective Evaluation}\label{sec:metrics}
As the actual goal of the study is to sort the candidate samples given an audio query $\mathbf{q}_i$, the primary evaluation metric is the $\text{rank}_i$ of the ground-truth samples for a given query (i.e., the samples that originate from the same song as the query). For that, we sort all drum samples in descending order according to the cosine similarity to a given encoded query and determine how early in the list a ground-truth sample is located. We also divide the $\text{rank}_i$ by $N$ list entries to obtain normalized ranks $\in (0, 1]$. As we can extract several queries ($4$-second-long random excerpts) for each song in the test set, and as there are several drum samples assigned to each song, we perform this test for every song in the test set $50$ times (resulting in $|Q| = 27$k evaluations) and average the resulting ranks. The explanation above results in the Mean Normalized Rank summarized as

\begin{equation}
R_{\text{mn}} = \frac{1}{|Q|} \sum_{i=1}^{|Q|} \frac{\text{rank}_i}{N}.
\end{equation}

In addition, we also report the \emph{Median} Normalized Rank $R_{\text{md}}$, as this is an indicator of how well the model performs \emph{most of the time}, ignoring possible outliers.

Besides also reporting the contrastive loss $\mathcal{X}$ (see Equation \ref{eq:loss}), we also evaluate how well the model has learned to cluster electronic and acoustic queries (i.e., mixtures) and keys (i.e., drum samples). To that end, we perform a binary (acoustic/electronic) $k$-nn classification (with $k=50$) on the encodings separately for both the queries and the keys. From this classification task, we report the likelihood ($L_q$ and $L_k$, respectively) of the data under the predictions.

In order to better understand the influence of different design choices, we perform ablation studies and report the metrics for each model and training variants. The different ablation scenarios are described in the following.

\begin{itemize}
\item{\texttt{2Enc} denotes our proposed setup with $2$ separate encoders instead of a shared encoder for both, queries and samples.}
\item{\texttt{PTrain} denotes a variant that starts training from an EfficientNet that was pre-trained on ImageNet, instead of from randomly initialized weights.}
\item{In \texttt{Aug}, data augmentation is used.}
\item{\texttt{VCReg}, is using the variance and co-variance regularization terms (see Equations \ref{eq:var} and \ref{eq:covar}).}
\item{In \texttt{SMix}, we randomly mix different numbers n of random stems to form a query (where $n>1$). If \texttt{SMix} is not used, we always mix all stems of a song.}
\item{For \texttt{QSInv}, in addition to sampling positive pairs from queries and corresponding samples, we also include as positive pairs two queries and two samples originating from the same song.}
\end{itemize}

\subsection{User Study}\label{sec:user_study_method}
\begin{table*}[t]
\centering
\footnotesize
\begin{tabular}{lllllllllll}
\toprule
& \multicolumn{5}{c}{Queries: full mixtures}  & \multicolumn{5}{c}{Queries: sparse mixtures}  \\

\cmidrule(r){2-6}
\cmidrule(r){7-11}
Variant & $\mathcal{X}$ & $R_{\text{mn}}$ & $R_{\text{md}}$ & $L_q$ & $L_k$ & $\mathcal{X}$ & $R_{\text{mn}}$ & $R_{\text{md}}$ & $L_q$ & $L_k$ \\
\midrule
\texttt{2Enc+PTrain+Aug+VCReg+SMix} & 3.614     & \textbf{0.105}  & \textbf{0.032} & 0.9940 &  0.9767 & \textbf{3.761}     & \textbf{0.124}  & 0.043 & \textbf{0.9905} &  0.9763 \\
\texttt{2Enc+PTrain+Aug+VCReg+SMix+QSInv} & 3.718    & 0.120  & 0.037 & 0.9900 & 0.9782 & 3.818     & 0.136  & 0.047 & 0.9862 &  0.9768 \\
\texttt{2Enc+PTrain+Aug+VCReg} & 4.001   & 0.124  & 0.061  & 0.9825 & 0.9732 & 4.389  & 0.183 & 0.100 &  0.9635 & 0.9724 \\
\texttt{2Enc+PTrain+VCReg+SMix} & 3.742    & 0.128  & 0.037 & 0.9945 & \textbf{0.9895} & 3.780     & 0.137  & \textbf{0.042} & 0.9901 &  \textbf{0.9898} \\
\texttt{2Enc+PTrain+Aug+SMix} & \textbf{3.575} & 0.116    & \textbf{0.032}  & \textbf{0.9946} & 0.9774 & 3.812     & 0.135  & 0.043 & 0.9893 &  0.9790 \\
\texttt{2Enc+Aug+VCReg+SMix} & 5.235 & 0.458  & 0.432  & 0.7268 & 0.7629 & 5.237     & 0.470 & 0.451 & 0.7188 & 0.7480 \\
\texttt{2Enc+Aug+VCReg+SMix+QSInv} & 4.174 & 0.164 & 0.079 & 0.9826 & 0.9566 & 4.387     & 0.181  & 0.091 & 0.9768 &  0.9585 \\
\texttt{PTrain+Aug+VCReg+SMix} & 3.853    & 0.121  & 0.047  & 0.9812 & 0.9809 & 4.000     & 0.137  & 0.058 & 0.9768 &  0.9819 \\
\texttt{2Enc+PTrain} & 3.883    & 0.140  & 0.053  & 0.9925 & 0.9795 & 4.399     & 0.205  & 0.089 & 0.9821 &  0.9803 \\
\bottomrule
\end{tabular}
\caption{Ablation study for different architectures and training scenarios tested on queries from full mixtures and queries from sparse mixtures (a sparse mixture is based on a random number $n$ of stems, where $n>1$). $\mathcal{X}$ denotes the contrastive loss and $R_{\text{mn}}$ and $R_{\text{md}}$ is the mean and median normalized rank, respectively, of ground truth samples. $L_q$ is the likelihood of the binary $k$-nn (electronic/acoustic) classification task for query, and $L_k$ for key encodings in the latent space.}
\label{tab:results}
\end{table*}

To evaluate the quality of the scoring function, we perform a user study in which we ask $10$ experts (with musical education or concerned with music production) to rate the quality of selected drum samples given a musical context. More precisely, we test their preference between drum samples that scored well according to our system and samples that are randomly picked from the dataset. The possible choices are to select one of the two proposed mixtures, or ``equal'' (if there is no preference for one of the choices), or ``skip'' (if the samples to be rated are not of the expected class or any other problem occurred). Every participant obtains $12$ single comparisons (presenting each of the $6$ percussion types twice) and provides $12$ ratings accordingly. After that, they can decide if they want to enter a further session (to provide $12$ further ratings). Participants are asked not to perform more than $4$ sessions to obtain a balanced result.

The procedure to obtain one comparison presented to a participant for rating is as follows. We pick a percussion type (e.g., snare) and a song from the test set that contains such a type as a single stem. We remove the stem containing that type, mix the remaining stems and pick ten $4$-second-long excerpts from randomly sampled positions of the resulting mixture. All these excerpts are then fed through the query encoder, and the mean is taken from the resulting encodings. The resulting mean vector acts as the query encoding for the current song. Using this query, the cosine similarities to the latent representations of all $63$k drum samples in the data set (obtained as described in Section \ref{sec:experiments}) are computed, and the ten samples with the \emph{highest similarities} are selected. Furthermore, ten drum samples of the corresponding percussion type are \emph{randomly picked} from the data set. The thereby obtained samples are used to generate 20 new versions of the current song (a version for each sample). 

To create a new version of a song, we perform an onset detection on the previously removed stem (containing the percussion type in question), position the selected samples in the estimated positions, and mix the resulting stem with the remaining stems (omitting the original stem of the percussion type in question). That way, we replace the original drum track with a track containing the drum sample to be evaluated. For a single user rating, we then contrast the mixtures of a \emph{randomly picked} and a \emph{highly rated} percussion sample and have participants indicate their preference between the two.

\subsection{Correlation Analysis}\label{sec:correlation_method}

In order to gain some insights into the latent space learned by the encoders, we perform a correlation analysis between audio features of neighboring data points. For that, different perceptual and spectral features are computed for the drum samples in the data set using the \emph{Audio Commons timbre models}\footnote{\url{https://github.com/AudioCommons/ac-audio-extractor}} for perceptual features (e.g., boominess, brightness, depth, hardness, roughness, warmth) and \emph{librosa}\footnote{\url{https://librosa.org/}} for spectral features (e.g., spectral centroid or spectral contrast). We also add an indicator if the respective drum sample originates from an electronic or acoustic song. In the former case, the ``electronic'' feature value is set to 1, in the latter case to 0.

The drum samples are mapped into the sample encoder's latent space, and $k$-means clustering is performed using $k=24$. In each cluster, the mean of each audio feature is used as a specific observation of that audio feature variable. Then, the Pearson correlation coefficient is computed between all such variables (separately for every percussion type). As a result, we can derive statements like ``whenever the loudness of a kick drum is high, the boominess of the snare tends to be low''. Note that for computing the audio features, we normalize every sample to $0.5$ seconds (i.e., cut or pad) so that the audio feature computation is not influenced by the length of the audio file (as some features are computed by averaging several spectrogram frames). To obtain the latent encodings used to compute the clusters, we use the regular $4$-second-long samples.

\section{Results and Discussion}\label{sec:results}

\begin{figure}
\vspace{-2.5mm}
\includegraphics[width=1.\linewidth]{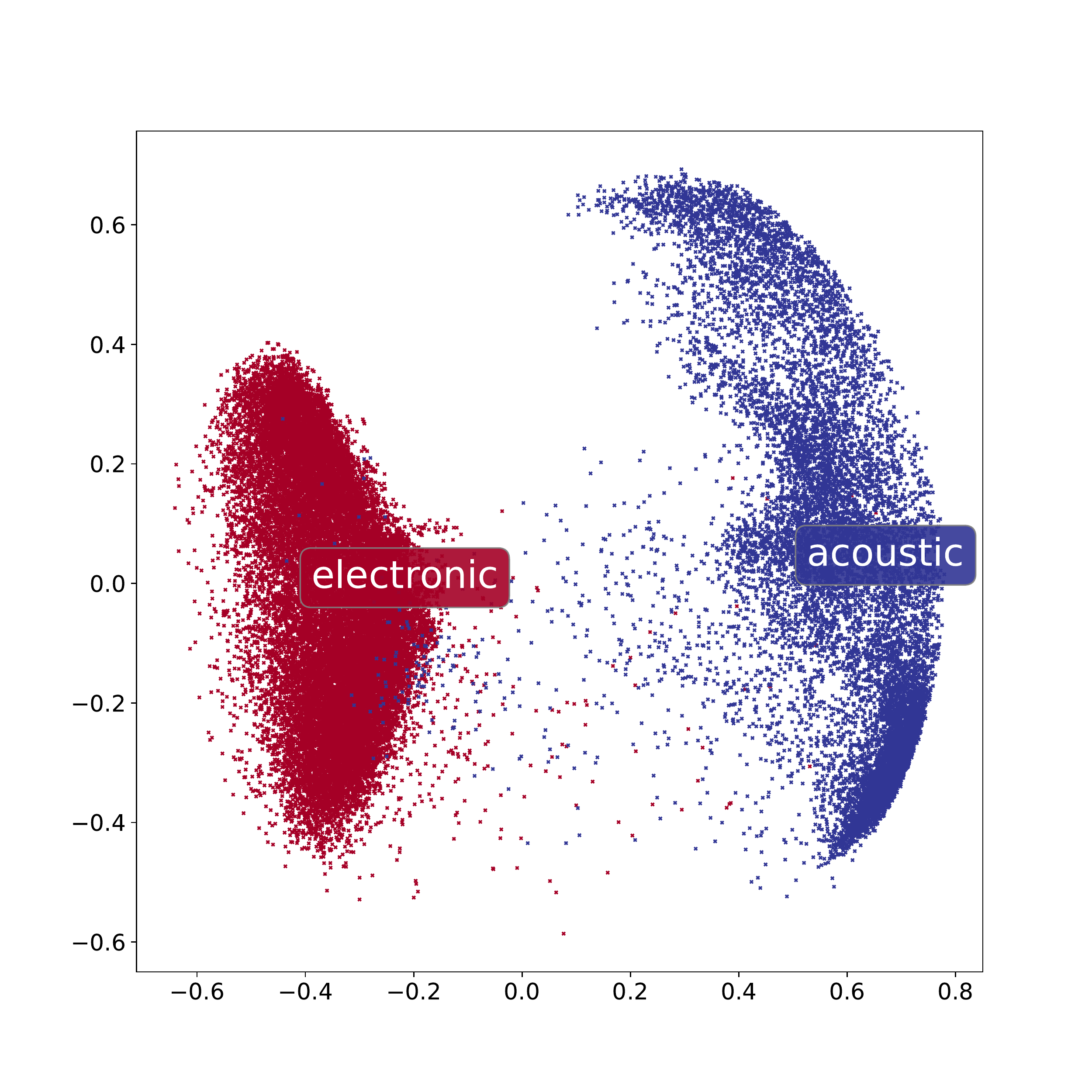} \\
\vspace{-2.5mm}
\caption{Principal Component Analysis (PCA) of drum sample encodings. Red dots indicate samples originating from electronic music and blue dots indicate samples originating from acoustic music.}
\label{fig:pca}
\end{figure}
\subsection{Objective Evaluation}\label{sec:objective}

In Table \ref{tab:results}, we report the contrastive loss, the Mean and Median Normalized Rank, and the query and sample electronic/acoustic classification likelihood for different architectures and training scenarios (as described in Section \ref{sec:metrics}). For each run, we train for 600 epochs and pick the saved checkpoint of the epoch with the best performance (regarding the Mean Normalized Rank $R_{\text{mn}}$) on the evaluation set for evaluating the test set. We report the results when evaluating the models with queries that are drawn from mixtures where all stems of a song are used (denoted as ``full mixtures'') and mixtures with a random number $n$ of stems (denoted as ``sparse mixtures'', where $n>1$). The latter scenario is crucial in music production, where sample selection may occur at an intermediate state of the project, when several instruments are still missing.

The results show that our proposed architecture and training procedure (first row in Table \ref{tab:results}) performs best in the mean and median rank metrics ($R_{\text{mn}}$ and $R_{\text{mn}}$) for both query scenarios (note that the random guessing baseline is $0.5$). It is somewhat surprising to us that \texttt{QSInv} (i.e., imposing additional invariance for queries/samples originating from the same song) does not improve but rather worsens the result (cf. row $2$ of the table). Apparently, the relaxation of the space caused by not using this regularization helps to perform the main objective.

We can see for \texttt{SMix} that it is vital to mix different (numbers of) stems on the fly during training (cf. third row in Table \ref{tab:results} where this has not been done). \texttt{SMix} does not only help for the  ``sparse mixtures'' scenario at test time (right part of the table) but also in the ``full mixtures'' scenario (left part of the table).

Using a pre-trained EfficientNet (\texttt{PTrain}) makes a considerable difference in the overall performance and training dynamics. In fact, when not using pre-trained weights, it is necessary to add \texttt{QSInv} in our experiments. Otherwise, the encoders do not learn at all (see row $6$ in Table \ref{tab:results}).

The last row of Table \ref{tab:results} presents the results of a scenario where all training optimizations (augmentation, variance, co-variance regularization, and sparse mixing) are turned off. It can be seen that the performance deteriorates substantially, particularly in the ``sparse mixtures'' test scenario.

Finally, the results of the classification likelihood (mostly around $99\%$) show that the models have implicitly learned to separate electronic and acoustic content (cf. Figure \ref{fig:pca}, showing a PCA of the drum sample encodings). Interestingly, this separation works consistently better if no data augmentation is used (see row $4$ in Table \ref{tab:results}). Electronic samples are generally ``cleaner'' than those extracted from acoustic music. Augmentation (e.g., adding noise) may remove some of these characteristics, making it harder to discriminate between the two classes.

\subsection{User Study}\label{sec:user_study}
\begin{figure}
\vspace{-2mm}
\includegraphics[width=1.\linewidth]{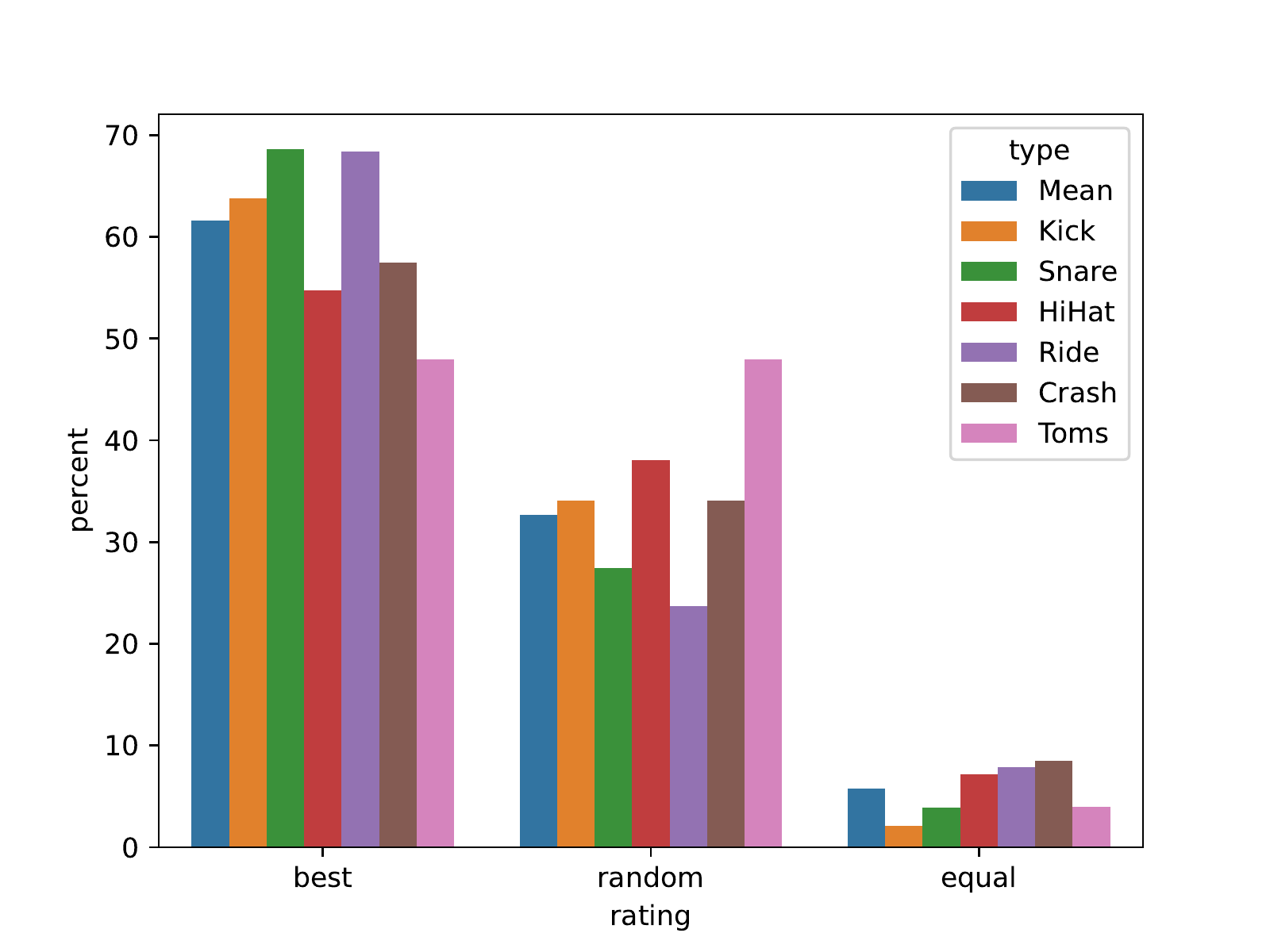} \\
\vspace{-3mm}
\caption{Preference ratings of participants in the user study, separated by percussion type (the blue bar ``Mean'' shows the mean of all ratings). ``best'' are mixtures with samples that \emph{scored highest} by our method, and ``random'' denotes mixtures with \emph{random samples} from the data set. An ``equal'' rating means no particular preference.}
\label{fig:userstudy}
\end{figure}
Figure \ref{fig:userstudy} shows the results of the listening test (based on $300$ ratings in total, omitting skipped ratings). On average (``Mean''), the samples that are ranked ``best'' by our method were preferred approximately twice as often as random samples ($61.57\%$ to $32.64\%$). It is also interesting that for most percussion types, the human preference for ``best'' is similar to or better than average (with Snare and Ride being the most salient), while human ratings disagree with our system's choices, particularly often for HiHat and Toms. A hypothesis why HiHats perform worse is that we merged both open and closed hi-hat in one group. When replacing the original hi-hats with the selected ones, there is an equal chance that an open hi-hat is used for a rhythm that is meant to be instantiated with a closed hi-hat and vice versa. This is particularly problematic in electronic music, where closed hi-hats are often used with minimal temporal intervals. The reason why Toms do not work well is probably due to less available training data (many tracks do not have tom stems). Examples of how the user study was presented can be found on the accompaniment website.$\,^\text{8}$

\subsection{Correlation Analysis}\label{sec:correlation_}
\begin{figure}
\includegraphics[width=1.\linewidth]{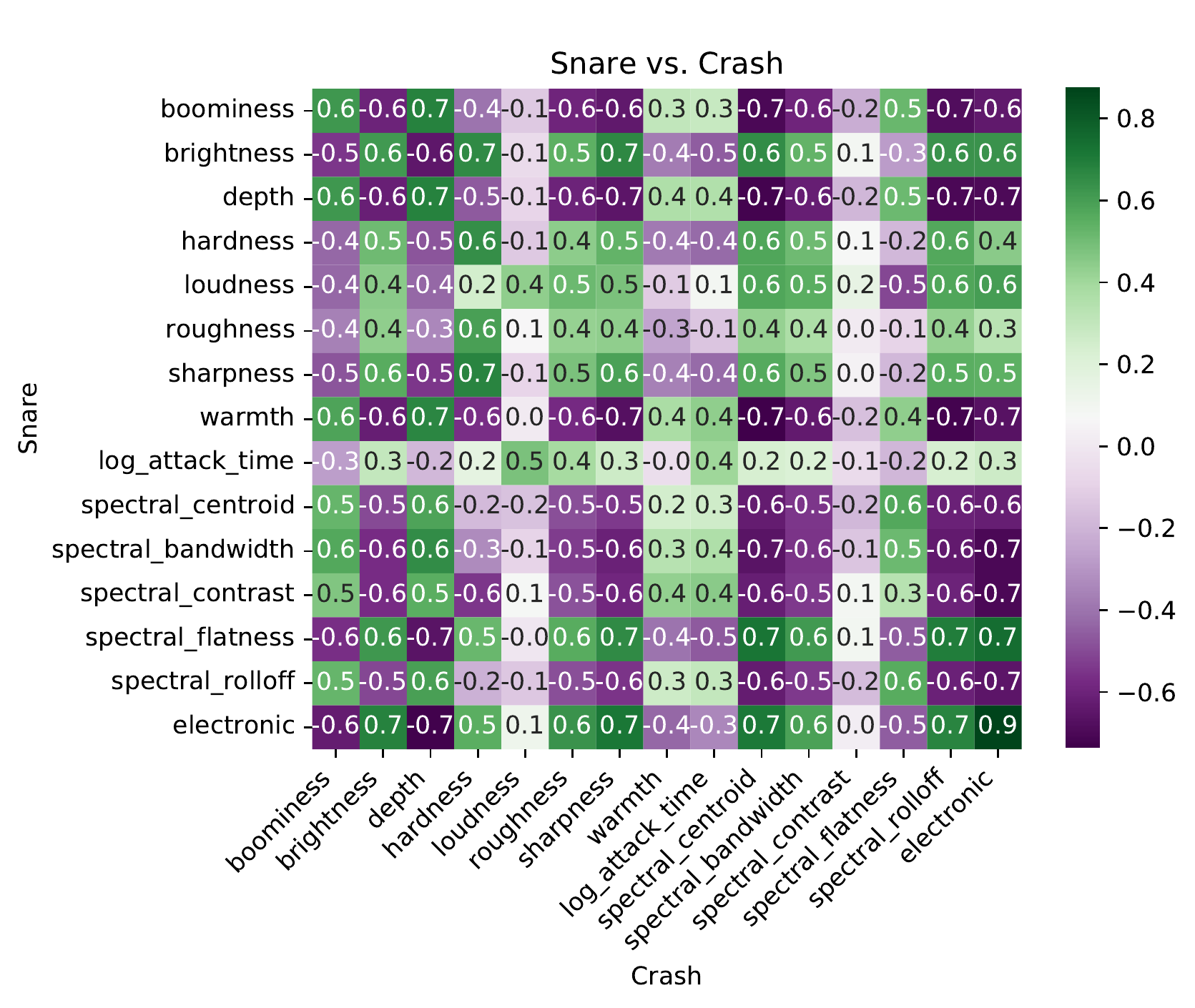} \\
\vspace{-2mm}
\caption{Correlations between perceptual and spectral features (and electronic / acoustic indicator) of Snare and Crash drum samples that are close in the latent space (i.e., scored to fit well in the same musical context).}
\label{fig:correlation}
\end{figure}

Figure \ref{fig:correlation} shows the correlation coefficients between audio features of snare and crash drum samples over clusters in the latent space (as described in Section \ref{sec:correlation_method}, note that the entirety of the percussion-type combinations is shown on the accompaniment website).\footnote{\url{https://sites.google.com/view/samplematch}}

To our knowledge, there is no theory about the aesthetic rules of drum sample selection. Therefore, this analysis is particularly informative. It is interesting to see that strong correlations exist at all between the audio features of snare and crash found in the same clusters of the latent space (as is true for most other percussion-type combinations). Consequently, the characteristics that make drum samples fit in the same musical contexts can (also) be explained by such lower-level features. Using such correlation coefficients makes it possible to understand how the learned space is organized. More importantly, it is also possible to derive rules and make the method explainable. For example, it is possible to extract statements like ``if the snare of a song sounds warm, the crash should not sound bright'' (because ``warmth'' of the snare and ``brightness'' of the crash are negatively correlated with a coefficient of $-0.6$).

When looking at the diagonal of the correlation matrix in Figure \ref{fig:correlation}, we see that the perceptual features tend to be positively correlated (from ``boominess'' to ``warmth''), while the spectral features tend to be negatively correlated (from ``log attack time'' to ``spectral rolloff''). Possibly, as snare and crash have similar perceptual characteristics, they should not get into each other's way regarding their frequency ranges. Whenever the snare occupies the higher frequencies, the crash occupies the lower frequencies and vice versa (according to the ``spectral centroid'' entry of $-0.6$ in the diagonal).

Finally, the correlations with the ``electronic'' indicator are also informative. It shows that a snare (and partly a crash) in electronic music tends to be more intense than in acoustic music, with more brightness, loudness, and sharpness. At the same time (looking at the spectral centroid), in electronic music, a snare tends to occupy rather lower frequencies than in acoustic music, while the opposite is true for crash (where the spectral centroid is positively correlated with electronic snares - having a correlation coefficient of $0.7$). The high correlation of ``electronic'' ($0.9$) in the diagonal shows that electronic snares and crashes tend to be in the same clusters of the latent space.

\section{Conclusion and Future Work}\label{sec:conclusion}
We introduced a method that automatically scores the fit of drum samples to a given musical context. As the method is thought to be used in a music production context, the audio queries used for training are based on ``sparse mixtures'', which allows scoring samples at different stages of the music production process. Results show that this form of data augmentation generally improves performance, also when queries are computed from complete mixtures at test time. Critically, the automatic system should agree with human judgment. Therefore, we performed a user study that tests this agreement. It could be shown that the drum samples that are highly rated by our system are also preferred by human experts approximately twice as often as randomly selected samples.

As the architecture and training procedure combine different ideas, we performed an ablation study, where it became clear that starting with a pre-trained model is crucial for a good final performance and that additional choices (like using variance and co-variance regularization, data augmentation, and on-the-fly sparse mixing) improve the results considerably.
It was also shown that audio features of drum samples whose encodings are close in the latent space are highly correlated, and we demonstrated that the observed correlations can also be interpreted to derive rules.

Our proposed method is not limited to drum samples but can potentially be used to retrieve different kinds of musical material based on learned aesthetic principles. Also, in the future, we want to scale up the system by using bigger data sets and a more generic pre-processing strategy. Currently, single drum samples are extracted before training to obtain a controlled experiment setup and application scenario. Developing a system that uses stems directly would apply better to audio sources of any type and would therefore greatly increase the applicability of our method in music production.

\section{Acknowledgements}
I want to thank the reviewers for their extraordinarily thorough suggestions to improve this work and my colleagues at Sony CSL that assisted me with technical expertise and emotional support.

\bibliography{bib_sl,bib_aa}
\end{document}